\documentclass[reprint, aps, onecolumn, amsmath, amssymb]{revtex4-2}

\usepackage[margin=2.6cm]{geometry}

\usepackage{hyperref}
\hypersetup{
    colorlinks = true,
    urlcolor   = blue,
    citecolor  = black,
}
\hypersetup{colorlinks,linkcolor={blue},citecolor={blue},urlcolor={blue}} 
\usepackage{graphicx}
\usepackage{dcolumn}
\usepackage{bm}
\usepackage[separate-uncertainty = true,multi-part-units=single]{siunitx}
\usepackage{mathtools}
\usepackage{gensymb}
\usepackage{amsmath}
\usepackage{physics}
\usepackage{stmaryrd}
\usepackage[nameinlink,noabbrev]{cleveref}
\crefformat{section}{\S#2#1#3}%
\crefformat{subsection}{\S#2#1#3}
\crefformat{subsubsection}{\S#2#1#3}
\crefformat{figure}{#2Fig.~#1#3}
\crefmultiformat{figure}{Figs.~#2#1#3}{ and~#2#1#3}{, #2#1#3}{ and~#2#1#3}
\usepackage{amssymb}
\usepackage[svgnames]{xcolor}

\begin{document}

\preprint{APS/123-QED}

\title{Periodic dynamics in viscous fingering\\}

\author{Jack Lawless}
\author{Andrew L. Hazel}%
\author{Anne Juel}
\email{anne.juel@manchester.ac.uk}
\affiliation{Manchester Centre for Nonlinear Dynamics, The University of Manchester, Oxford Road, Manchester, M13 9PL, UK\\}

\date{\today}

\begin{abstract}
The displacement of a viscous liquid by air in the narrow gap between two parallel plates -- a Hele-Shaw channel -- is an exemplar of complex pattern formation. Typically, bubbles or fingers of air propagate steadily at low values of the driving parameter. However, as the driving parameter increases, they can exhibit disordered pattern-forming dynamics. In this paper, we demonstrate experimentally that a remote perturbation of the bubble's tip can drive time-periodic bubble propagation: a fundamental building block of complex unsteady dynamics. We exploit the propensity of a group of bubbles to self-organise into a fixed spatial arrangement in a Hele-Shaw channel with a centralised depth-reduction in order to apply a sustained perturbation to a bubble's shape as it propagates. We find that the bubble with a perturbed shape begins to oscillate after the system undergoes a supercritical Hopf bifurcation upon variation of the tip perturbation and dimensionless flow rate. The oscillation cycle features the splitting of the bubble's tip and advection of the resulting finger-like protrusion along the bubble's length until it is absorbed by the bubble's advancing rear. The restoral of the bubble's tip follows naturally because the system is driven by a fixed flow rate and the perturbed bubble is attracted to the weakly unstable, steadily propagating state that is set by the ratio of imposed viscous and capillary forces. Our results suggest a generic mechanism for time-periodic dynamics of propagating curved fronts subject to a steady shape perturbation.
\end{abstract}

\maketitle


\section{Introduction}

\begin{figure}
\includegraphics[width=\textwidth, clip]{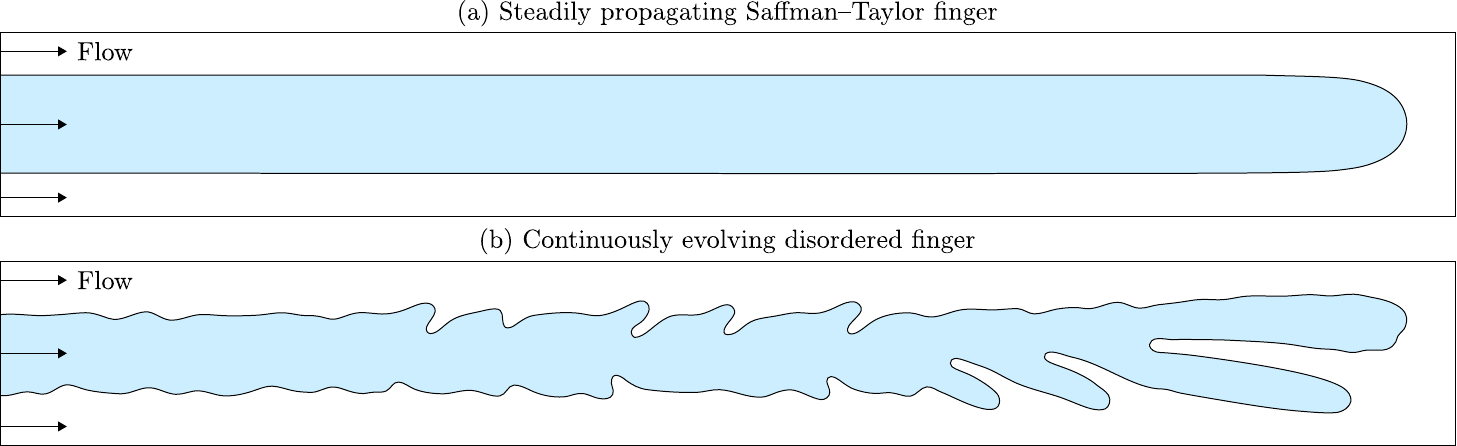}
\caption{The typical fingering patterns that develop when a constant flux of air invades a viscous liquid-filled Hele-Shaw at two different values of the dimensionless parameter $1/B = 12 \: \alpha^2 \: \mathrm{Ca}$, where $\alpha = W^* / H^*$ is the channel's cross-sectional aspect ratio and $\mathrm{Ca} = \mu V^* / \sigma$ is a capillary number. Here, $\mu$ is the dynamic viscosity of the liquid, $\sigma$ is the surface tension and $V$* is the finger's propagation speed. (a) The single, steadily propagating and symmetric Saffman--Taylor finger of air that develops for $1/B = 1200$. The channel's aspect ratio is $\alpha = W^*/H^* = 40$ and the liquid's dynamic viscosity is $\mu =$ \SI{0.019}{\pascal \second}. (b) The disordered finger of air that develops for $1/B = 15000$. The channel's aspect ratio is $\alpha = W^*/H^* = 60$ and the liquid's dynamic viscosity is $\mu =$ \SI{0.095}{\pascal \second}. The surface tension is $\sigma =$ \SI{21}{\milli \newton \per \metre} in both cases.}
\label{fig:ST_vs_disordered}
\end{figure}

Viscous fingering is an archetype of complex pattern formation \citep{perspectives}. However, under certain conditions, this phenomenon can also lead to simple long-term behaviours. For example, in a rectangular Hele-Shaw channel of width $W^*$ that is much larger than its depth $H^*$, an initially planar air-liquid interface is linearly unstable to infinitesimal perturbations when it is driven by a constant volume flux flow. The interface develops a series of competing finger-like protrusions; those that forge ahead tend to shield their neighbours until a single, steadily propagating and symmetric finger of air prevails --- the so-called "Saffman--Taylor finger" (figure~\ref{fig:ST_vs_disordered}a) \citep{Saffman1958}. If $H^*$ is sufficiently small relative to $W^*$, gravitational, inertial and three-dimensional effects can be neglected and the resulting equations of motion are described in terms of a single dimensionless parameter $1/B = 12 \: \alpha^2 \: \mathrm{Ca}$, where $\alpha = W^*/H^*$ is the channel's cross-sectional aspect ratio and $\mathrm{Ca} = \mu V^* / \sigma$ is a capillary number \citep{mclean1981}. Here, $\mu$ is the dynamic viscosity of the liquid, $\sigma$ is the surface tension and $V^*$ is the finger's propagation speed. The Saffman--Taylor finger was found to be linearly stable to infinitesimal perturbations for all values of $1/B$ in the McLean--Saffman depth-averaged lubrication model \citep{Kessler}. However, the finger destabilises at a finite value of $1/B$ in experiments and deposits a disordered, continuously evolving fingering pattern through repeated tip-splitting events and subsequent finger competition as it propagates (figure~\ref{fig:ST_vs_disordered}b) \citep{park_homsy}. The critical value of $1/B$ that is required for the onset of the instability is not well-defined because it exhibits an exponentially sensitive dependence on the level of background noise that is present in the system \citep{bensimon}. For example, the critical value of $1/B$ typically exceeds $10^4$ in smooth-walled channels that are filled with a purified liquid \citep{tabeling}. However, this can be reduced ten-fold by suspending small particles in the liquid in order to systematically increase the level of background noise \citep{chevalier}. The physical mechanism that underlies this phenomenon is yet to be established. However, it bears the hallmarks of a subcritical (i.e perturbation-driven) transition to disorder and this motivates an exploration of the system's nonlinear dynamics. 

The subcritical transition from laminar flow to turbulence in linearly stable wall-bounded shear flows is probably the most renowned example of a perturbation-driven transition to disorder in the context of fluid mechanics~\citep{turbulence_review} and, despite its complexity, significant progress has been made in understanding this phenomenon by using dynamical systems theory as a conceptual framework \citep{eckhardt}. The cornerstone of this particular approach is to interpret a system's temporal evolution as an exploration of the invariant objects (e.g. equilibria and periodic orbits) that are contained in its phase space. The system's trajectories are either attracted or repelled by these objects and this can lead to complex transient dynamics. There has been a particular emphasis placed on the identification of weakly unstable periodic orbits in studies of the transition to turbulence because of their role in facilitating complex time-evolving dynamics \citep{duguet}. The turbulent trajectories typically meander between a multitude of weakly unstable periodic orbits because they are attracted to their stable manifolds as the system evolves in time; this behaviour suggests that such orbits are embedded in high-dimensional chaotic structures and act as the fundamental building blocks of turbulent dynamics \citep{Kawahara, budanur, Cvitanovic}. We hypothesise that weakly unstable periodic orbits could play a similar role in the transition from steady to disordered front propagation and, thus, underlie the formation of complex patterns in viscous fingering.


However, there is no conclusive evidence to date that viscous fingering in a Hele-Shaw channel of uniform depth supports time-periodic dynamics. The absence of inertial forces in our system means that nonlinearities arise exclusively at the air-liquid interface and, thus, the dynamical state of an air finger is encapsulated in its shape. The family of steady, symmetric and single-tipped fingers are the only stable modes of propagation that exist in a Hele-Shaw channel of uniform depth. However, numerical simulations of a depth-averaged lubrication model have uncovered several weakly unstable families of steadily propagating fingers that exist alongside it \citep{Saffman1959, romero, vanden-broeck, FrancoGomez2016}. On the other hand, weakly unstable periodic modes of propagation have not been identified and this is not entirely surprising because time-dependent states are, in general, more challenging to find. We note that seemingly periodic tip-splitting was reported prior to the onset of disordered dynamics in experiments \citep{park_homsy}. However, the practical limitations of the experimental setup rendered it unclear if this was in fact a periodic mode of propagation or, instead, complex transient behaviour.

In this paper, we demonstrate a generic mechanism that supports periodic dynamics in finger propagation. We take inspiration from the periodic dynamics that have previously been reported upon modifying the environment in which the air finger propagates. For example, the curvature of the finger's tip can be modified by forcing it to aggregate with a small bubble and this leads to a stable meandering of the finger's tip about the channel's centreline as it propagates \citep{couder}. Unsteady modes of finger propagation have also been found in benchtop models of airway reopening, in which the upper boundary of the channel is replaced by a deformable elastic sheet. The channel is filled with a viscous liquid, initialised in a partially collapsed state and subsequently reopened by injecting an air finger into it at a constant flow rate. The injected air works against viscous and capillary forces in the liquid and, in addition, the sheet's elastic resistance to deformation. For small levels of initial collapse, the finger propagates steadily and adopts a shape that is broadly similar to that it would assume in an equivalent rigid-walled channel \citep{ducloue}. However, a variety of unsteady propagation modes were observed in highly collapsed channels at large values of the capillary number in experiments \citep{cuttle} and numerical simulations of a fully coupled depth-averaged lubrication model were used to identify several periodic modes of finger propagation \citep{Fontana_2023}. Periodic dynamics have also been found in simpler geometries, such as in a channel whose cross-section is partially occluded by an axially-centred depth-reduction, or "rail". The rail-induced reduction in the channel's depth modifies the transverse curvature of the air-liquid interface and this leads to stable oscillations of the finger's tip in experiments \citep{Pailha, Jisiou} and numerical simulations \citep{Thompson2014}. However, a restoring mechanism was not identified. Finally, the displacement of a liquid by a finite bubble is an analogous problem to semi-infinite finger propagation and, in a rail-occluded channel, steadily propagating bubbles were found to exhibit unpredictable long-term evolutions when they were perturbed by a localised channel constriction \citep{Lawless}. The depth-averaged lubrication model was used to attribute this behaviour to a sensitive dependence on initial conditions in the vicinity of a weakly unstable periodic orbit but the physical mechanism underlying periodic dynamics was not explored.

Here, we focus on the propagation of a finite bubble which is remotely perturbed; this can be achieved in a rail-occluded Hele-Shaw channel because groups of bubbles are able to self-organise into a multitude of stable spatial arrangements. This behaviour contrasts to that of bubbles in a Hele-Shaw channel of uniform depth because pairs of neighbouring bubbles will always separate or aggregate depending on their ordering. The constituent bubbles of a stable arrangement --- henceforth referred to as a "formation" --- are arranged in alternation on opposite sides of the rail and remain separated from one another because they translate steadily at the same speed. The stable formations that were described in \cite{formations} consisted of bubbles that propagated steadily with fixed shapes. However, in this paper, we demonstrate that an individual bubble can, in fact, exhibit stable periodic dynamics when it is propagated as part of a stable formation in an isolated region of the system's parameter space.

The paper is organised as follows. The experimental set-up and the procedures that are used in order to initialise a bubble of prescribed volume and vary its initial position are described in section~\ref{sec:setup}. The fundamental features of a group of bubbles propagating as part of a stable formation are outlined in section~\ref{sec:formations}. We provide a general overview of the oscillatory dynamics that arise in our system in section~\ref{sec:periodic}. The underlying physical mechanism that gives rise to oscillatory dynamics is explained by analysing the behaviour of the bubble's front during a single oscillatory cycle in sections~\ref{sec:splitting} and \ref{sec:restoral}. We conclude by discussing the significance of our results for the existence of a generic mechanism for periodic dynamics in viscous fingering in section~\ref{conclusion}.

\begin{figure}
\includegraphics[width=\textwidth, clip]{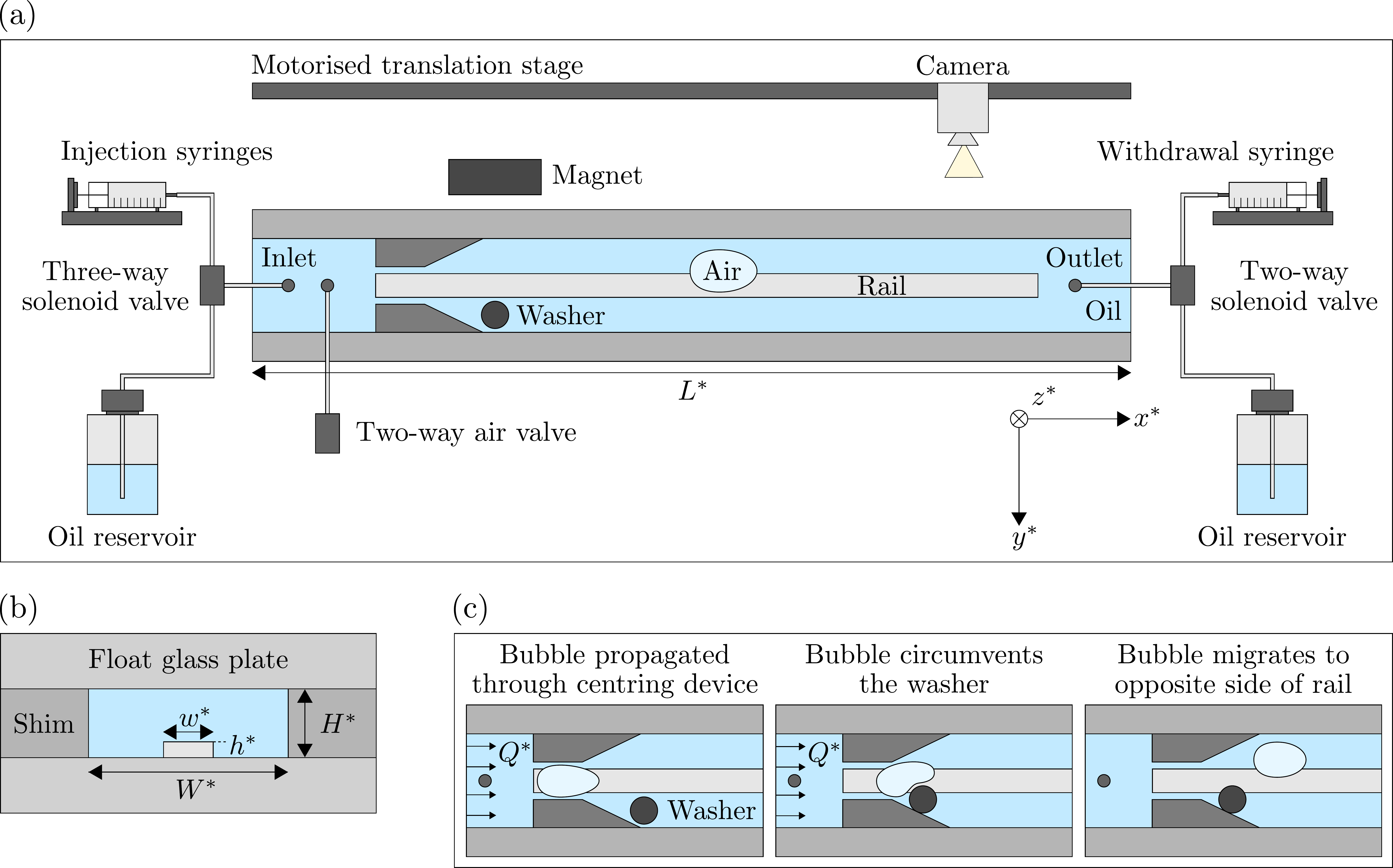}
\caption{(a) The horizontally levelled, depth-perturbed Hele-Shaw channel. The channel inlet is connected to a network of syringe pumps and an external oil reservoir by a three-way solenoid valve. The channel outlet is connected to a syringe pump and an external oil reservoir by a two-way solenoid valve. The experiments are recorded in top-view by a steadily translating CMOS camera that is mounted onto a motorised translation stage. (b) The channel's cross-section. (c) The procedure that is used to initialise a bubble on a particular side of the rail; the washer's position is controlled non-invasively by a N52-grade neodymium magnet that is situated underneath the lower glass plate.}
\label{fig:setup}
\end{figure} 

\maketitle

\section{Experimental methods}
\subsection{Experimental set-up}
\label{sec:setup}
The experiments were performed in a horizontal Hele-Shaw channel that was partially occluded by an axially-centred rail (figure~\ref{fig:setup}a). The channel consisted of two float glass plates that were separated by two parallel sheets of steel shim. The width of the channel was $W^* =$ \SI{40.0 (1)}{\milli\metre} and its depth was $H^* =$ \SI{1.00 (1)}{\milli\metre}. The rail consisted of a thin strip of translucent PET tape of width $w^* = $ \SI{10.0 (1)}{\milli\metre} and thickness $h^* =$ \SI{24 (1)}{\micro\metre} that was applied along the centreline of the lower glass plate (figure~\ref{fig:setup}b). The channel was filled with silicone oil (Basildon Chemicals Ltd) of dynamic viscosity $\mu =$ \SI{0.019}{\pascal \second}, density $\rho =$ \SI{951}{\kg \per \cubic \metre} and surface tension $\sigma =$ \SI{21}{\milli \newton \per \metre} at the ambient laboratory temperature of \SI{21(1)}{\celsius}. The flow of oil was controlled by a network of syringe pumps and two solenoid valves; a three-way solenoid valve was used to connect the injection syringe pumps, an external oil reservoir that was held at atmospheric pressure and the channel inlet, whilst a two-way solenoid valve was used to connect the external oil reservoir, a withdrawal syringe pump and the channel outlet. Thus, depending on the configuration of the two solenoid valves, oil could either be injected into the channel at a constant volumetric flow rate $Q^*$ or withdrawn from the external oil reservoir in order to refill the injection syringes. The syringe pumps and solenoid valves were controlled by a custom-built LabVIEW script.

\begin{figure}
\includegraphics[width=\textwidth, clip]{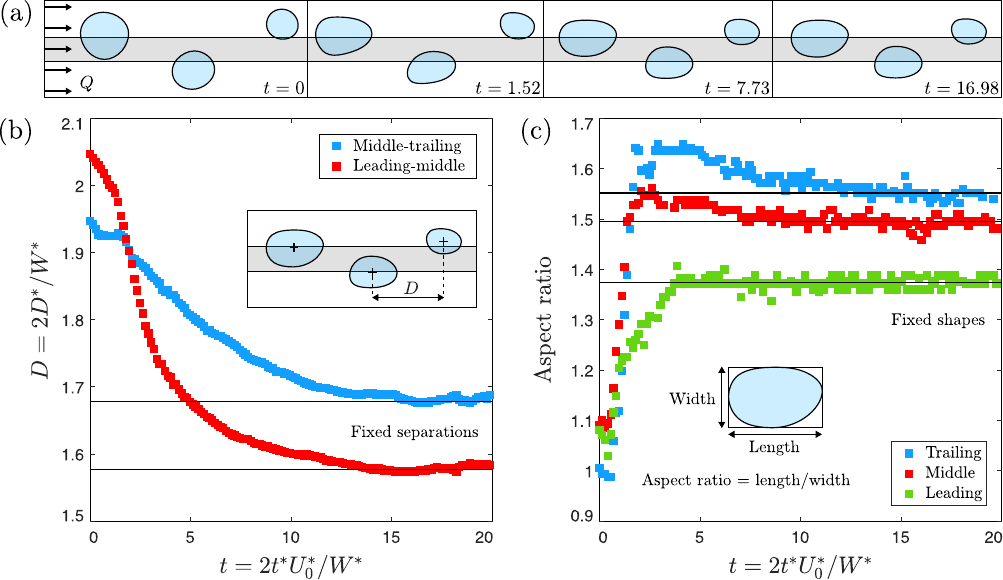}
\caption{(a) An example time-sequence of three bubbles organising into a stable formation in a rail-occluded Hele-Shaw channel when propagated from rest in an alternating initial configuration at $t=0$. The rail is indicated by the central grey-coloured band. The dimensionless bubble sizes are $r_1=0.30$, $r_2=0.40$ and $r_3=0.47$. The dimensionless flow rate (imposed from left-to-right) is $Q= \mu U_0^* / \sigma = 0.04$. (b) Time-evolution of the the dimensionless centroid separation $D = 2D^* / W^*$ between pairs of consecutive bubbles. (c) Time-evolution of the bubble aspect ratios.}
\label{fig:formation}
\end{figure} 

We generated three bubbles inside the channel by withdrawing prescribed volumes of oil whilst the two-way air valve (figure~\ref{fig:setup}a) was open to the atmosphere. The bubbles were propagated through a centring device, which consisted of a symmetric constriction and a linear expansion region, before being guided to their prescribed initial positions. We used a magnetic washer in order to asymmetrically constrict the channel and guide the bubbles to their prescribed initial positions. The washer's position was controlled non-invasively by an N52-grade neodymium magnet that was situated underneath the lower glass plate (figure~\ref{fig:setup}c) and the three bubbles were initialised on alternating sides of the rail by varying the side of the channel from which the constriction was applied.

The bubbles were propagated by injecting oil into the channel at a constant volumetric flow rate $Q^*$ and filmed in top-view by an overhead CMOS camera that was mounted onto a motorised translation stage. The camera was programmed to translate at a constant speed, which was selected by using an empirical relationship between the imposed flow rate and a bubble's speed, in order to ensure that the group of bubbles remained within its field of view throughout the experiment \citep{LifeAndFate}. The channel was illuminated from below by an LED light box in order to enhance the contrast of the air-fluid interface and the bubble contours were identified with a Canny edge-detection algorithm. We adopt the channel's half-width $W^*/2$ and the mean speed of oil in an unoccluded (i.e. rail-less) channel with equivalent cross-sectional dimensions $U_0^* = Q^* / W^* H^*$, respectively, as our characteristic length and velocity scales. The imposed flow rate is parametrised in terms of a dimensionless flow rate $Q = \mu U_0^* / \sigma$, which is a capillary number that is based on the mean speed of oil. The bubble's speed $U^*$ was determined by calculating the streamwise displacement of its centre of mass across a series of frames and the dimensionless bubble speed $U = U^* / U_0^*$ measures it relative to the mean speed of oil. The bubble's in-plane area $A^*$ was determined from image analysis and we parametrise its size in terms of a dimensionless radius $r = 2 \sqrt{A^* / \pi} / W^*$. 

\subsection{Stable bubble formations}
\label{sec:formations}

The physical conditions for periodic bubble dynamics were achieved by constructing a stable formation (i.e. a fixed spatial arrangement) of three bubbles propagating at a constant speed. The underlying physical mechanism that allows a group of bubbles to propagate in a stable formation in a rail-occluded Hele-Shaw channel is already described in detail in \cite{formations} and, thus, we will only outline the fundamental features. The three bubbles in the first panel of figure~\ref{fig:formation}a have been initialised in order of decreasing size in the direction of inlet-to-outlet on alternating sides of the rail. The leading bubble's size is $r_1=0.30$, the middle bubble's size is $r_2=0.40$ and the trailing bubble's size is $r_3=0.47$. The three bubbles are propagated from rest by imposing a constant dimensionless flow rate $Q = 0.04$ of the surrounding oil at $t=0$ and this causes them to deform. The separation between neighbouring bubbles initially decreases (figure~\ref{fig:formation}b) because larger bubbles tend to propagate faster than smaller bubbles \citep{LifeAndFate}. However, the middle and trailing bubbles slow down by adjusting their shape as they approach their preceding neighbours until they match that of the leading bubble. The three bubbles henceforth retain this particular spatial arrangement and propagate steadily with fixed shapes (figure~\ref{fig:formation}c). The stable formation is characteristically led by the smallest of its constituent bubbles and, importantly, this would be the slowest of the bubbles if they were all propagating in isolation because the propagation speed of an isolated bubble increases monotonically with its size. The leading bubble's motion is uninfluenced by the trailing bubbles; it adopts a near-identical shape to that it would assume in isolation and thereby sets the propagation speed of the stable formation. The trailing bubbles align themselves in the perturbation fields of their preceding neighbours and change their shapes by varying their respective overlaps of the rail in order to reduce their speeds to that of the leading bubble. In general, the extent of a particular bubble's shape perturbation increases as the difference between its size and that of the leading bubble increases due to the growing contrast in their isolated speeds.

\begin{figure}[t]
\includegraphics[width=\textwidth, clip]{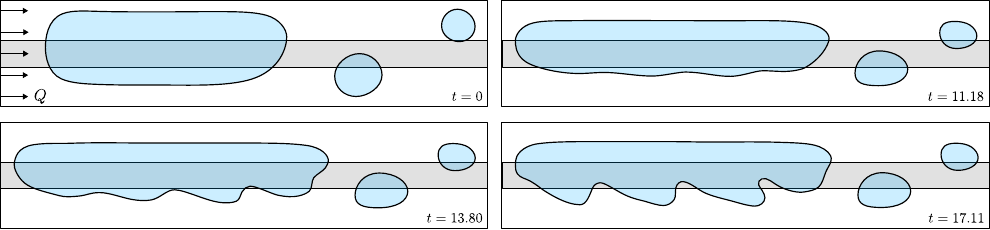}
\caption{Experimental time-sequence of a stable periodic bubble formation in a rail-occluded Hele-Shaw channel. The bubbles are initialised in an alternating initial configuration in the first panel and their sizes decrease in the direction of inlet-to-outlet (left-to-right). The rail is represented by the central grey-coloured band. The bubble sizes are $r_1 = 0.30$, $r_2 = 0.40$ and $r_3 = 1.30$ and the dimensionless flow rate is $Q= \mu U_0^* / \sigma = 0.04$.}
\label{fig:three_bubble_oscs_Q106}
\end{figure}

\section{Results:\protect\\}
\label{sec:results}

\subsection{Periodic bubble dynamics}
\label{sec:periodic}

\begin{figure}
\includegraphics[width=\textwidth, clip]{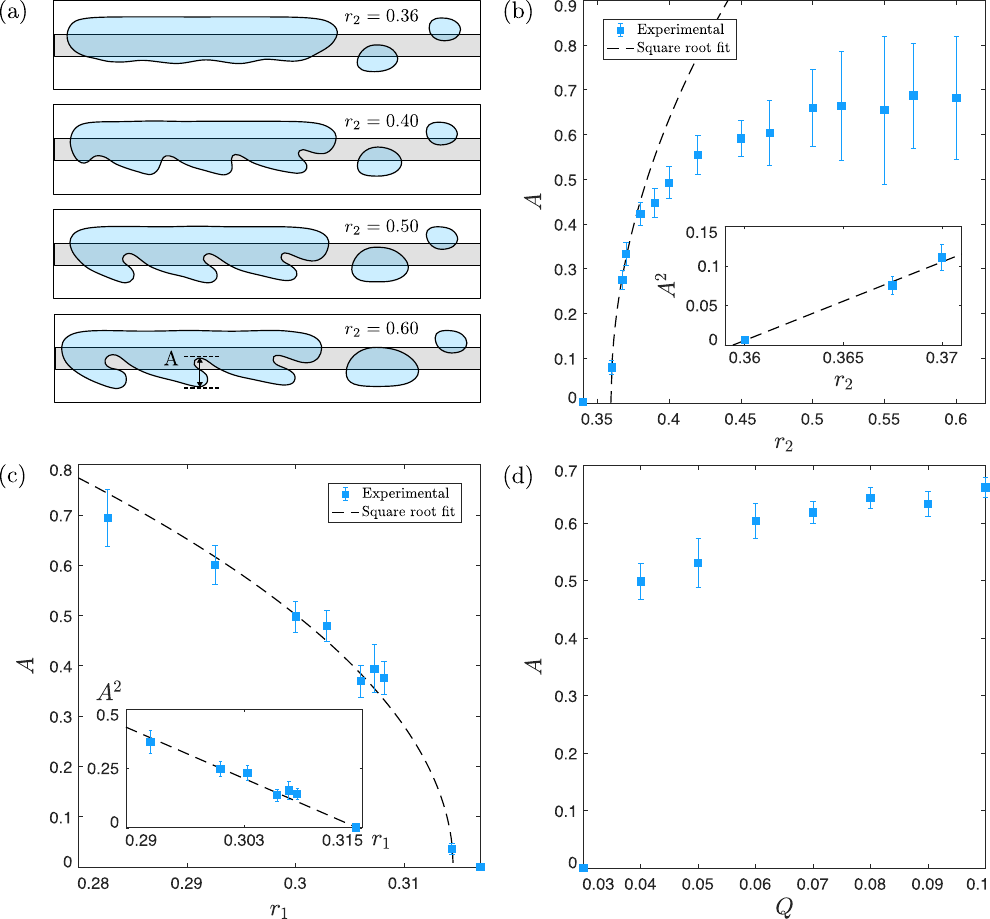}
\caption{(a) Snapshots of the quasiperiodic patterns that are deposited by the trailing bubble as the middle bubble's size $r_2$ is varied. (b--d) Variation of the quasiperiodic pattern's dimensionless amplitude $A = 2A^* / W^*$ as functions of (b) the middle bubble's size $r_2$, (c) the leading bubble's size $r_1$ and (d) the dimensionless flow rate $Q = \mu U_0^* / \sigma$. The error bars are the standard deviation of the data sets. The non-varying parameters are fixed at $r_1 = 0.30$, $r_2=0.40$, $r_3 = 1.30$ and $Q = 0.04$.}
\label{fig:amplitude}
\end{figure} 

\begin{figure}
\includegraphics[width=\textwidth, clip]{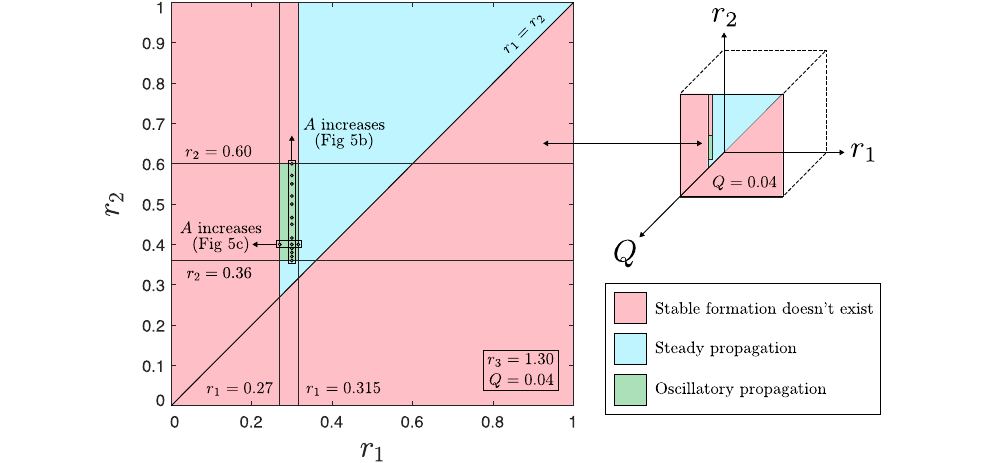}
\caption{Two-dimensional phase diagram of the trailing bubble's long-term behaviour as the leading bubble's size $r_1$ (horizontal axis) and middle bubble's size $r_2$ (vertical axis) are varied at a fixed value of the dimensionless flow rate $Q = \mu U_0^* / \sigma = 0.04$. The trailing bubble's size $r_3 = 1.30$ is constant. The background colours are used to delineate different long-term behaviours.}
\label{fig:phase_diagram}
\end{figure}

We fix the size of the trailing bubble $r_3=1.30$ such that it is large enough to propagate like a semi-infinite air finger \citep{Lawless} and, hence, there are three parameters that can be systematically varied: the leading bubble's size $r_1$, the middle bubble's size $r_2$ and the dimensionless flow rate $Q = \mu U_0^* / \sigma$. The trailing bubble's front oscillates (i.e. deforms periodically) as it propagates for particular combinations $(r_1, \: r_2, \: Q)$ of the system's parameters and the time-sequence in figure~\ref{fig:three_bubble_oscs_Q106} shows a typical example of this behaviour. The bubbles have been initialised in order of decreasing size in the direction of inlet-to-outlet on alternating sides of the rail in the first panel; the leading bubble's size is $r_1 = 0.30$, the middle bubble's size is $r_2 = 0.40$ and the dimensionless flow rate is $Q = 0.04$. The bubbles organise into their stable formation following the imposition of flow and, in a similar fashion to the stable formation of bubbles that was presented in figure~\ref{fig:formation}, the leading and middle bubbles propagate with fixed shapes. However, the trailing bubble does not propagate with a fixed shape because its front deforms periodically and this causes a quasiperiodic pattern of finger-like protrusions to develop along its underside in the final panel.

The trailing bubble oscillates because the system undergoes a supercritical Hopf bifurcation and a stable limit cycle emanates from an unstable steadily propagating multi-bubble state at particular combinations of its parameters. We find that the system can bifurcate following variation of one of its parameters whilst the other two are fixed at particular values. For example, whilst the leading bubble's size $r_1 = 0.30$ and dimensionless flow rate $Q = 0.04$ are fixed, the bifurcation occurs as the middle bubble's size is increased beyond the critical value $r_2 = r_{2\mathrm{c}} \equiv 0.36$. The quasiperiodic pattern that develops along the trailing bubble's underside consists of small-wave like disturbances in the immediate vicinity of the bifurcation point (e.g. first panel of figure~\ref{fig:amplitude}a) and develops into pronounced finger-like protrusions (e.g. fourth panel of figure~\ref{fig:amplitude}a) as the middle bubble's size is increased beyond the critical value. We use the lateral separation between the neighbouring peaks and troughs of the deposited pattern as the quantitative measure of its dimensionless amplitude $A = 2 A^* / W^*$. The pattern's amplitude increases monotonically from zero as $r_2$ exceeds $r_{2\mathrm{c}}$ (figure~\ref{fig:amplitude}b) and, in accordance with the local dynamical behaviour of a supercritical Hopf bifurcation, its initial increase in the vicinity of $r_{2\mathrm{c}}$ is directly proportional to $\sqrt{r_2-r_{2\mathrm{c}}}$ (inset of figure~\ref{fig:amplitude}b). The pattern's amplitude plateaus as the middle bubble's size is increased significantly beyond the critical value. Furthermore, the trailing bubble splits into two for $r_2 > 0.60$.


We find qualitatively similar behaviour by either decreasing the leading bubble's size (figure~\ref{fig:amplitude}c) or increasing the dimensionless flow rate (figure~\ref{fig:amplitude}d) whilst the other two system parameters are fixed at particular values and this suggests that there is a single supercritical Hopf bifurcation locus in the three-dimensional $(r_1, \: r_2, \: Q)$ parameter space. However, we have not sought to obtain a functional expression for this locus. 

The phase diagram in figure~\ref{fig:phase_diagram} classifies the trailing bubble's long-term behaviour in a two-dimensional slice $(r_1, \: r_2, \: Q = 0.04)$ of the three-dimensional parameter space. The red-coloured region indicates that a stable formation does not exist. There are three different reasons as to why a stable formation does not exist. Firstly, the leading bubble must be the smallest of the bubbles and, thus, a stable formation does not exist for $r_1 > r_2$. Furthermore, the leading bubble's size must be sufficiently large ($r_1 > 0.27$) because there is a maximum reduction in the trailing bubble's speed that can be facilitated by the rail \citep{Lawless}. Finally, the trailing bubble splits into two if the middle bubble's size is too large ($r_2 > 0.60$). The cyan-coloured region indicates that a stable formation does exist and the trailing bubble propagates with a steady shape. The tip perturbation that is imposed in this region is not sufficiently large to result in oscillations of the trailing bubble's front. The isolated green-coloured region ($0.27 \leq r_1 \leq 0.315$ and $0.36 \leq r_2 \leq 0.60$) indicates that a stable formation does exist and the trailing bubble oscillates. We have performed more experiments and our results suggest that the qualitative features of the phase diagram are similar at higher flow rates. Thus, having presented the general oscillatory dynamics that arise in this system, we now proceed to explore the behaviour of the trailing bubble's front during a single oscillatory cycle.

\begin{figure}[b]
\includegraphics[width=\textwidth, clip]{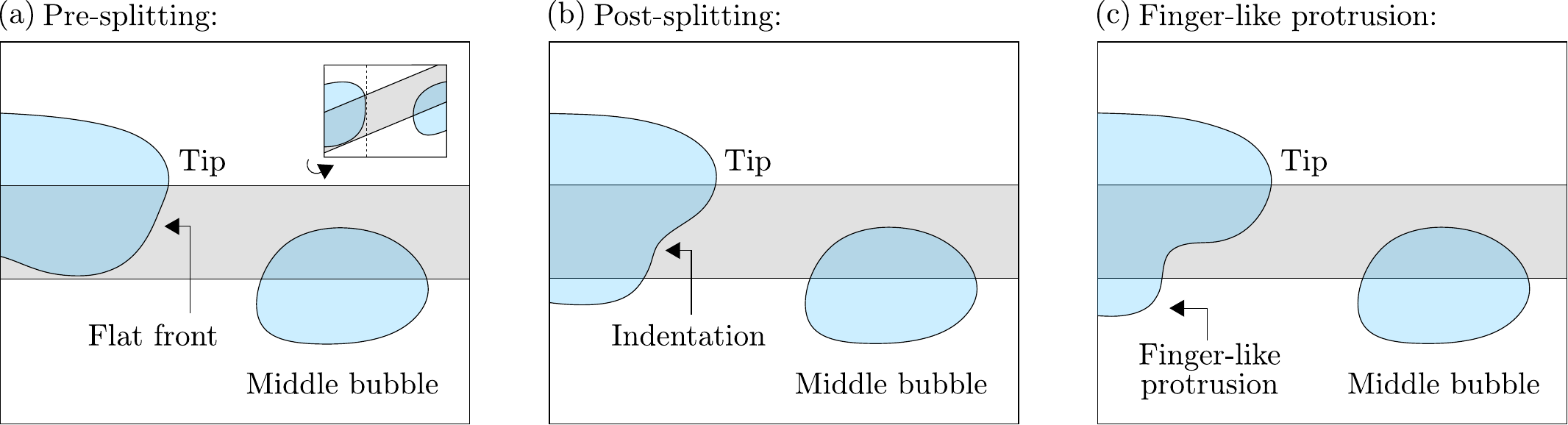}
\caption{The generic splitting dynamics of the trailing bubble's front. (a) The trailing bubble's flat front prior to splitting. (b) The indentation that develops underneath the trailing bubble's tip. (c) The finger-like protrusion that develops underneath the trailing bubble's tip. The bubble sizes are $r_1 = 0.30$, $r_2=0.40$ and $r_3=1.30$ and the dimensionless flow rate is $Q= \mu U_0^* / \sigma = 0.04$.}
\label{fig:flat_front}
\end{figure}

\subsection{Splitting dynamics and the advection of disturbances}
\label{sec:splitting}

The splitting of the trailing bubble's front follows a generic mechanism irrespective of the system's parameters; its onset occurs when the region of the trailing bubble's front that faces the middle bubbles rear is flat (figure~\ref{fig:flat_front}a). The flat region of the trailing bubble's front develops an indentation (i.e. a region of oppositely signed curvature) underneath its tip (figure~\ref{fig:flat_front}b). The onset of splitting is followed by the transient growth of a finger-like protrusion that develops underneath the indentation (figure~\ref{fig:flat_front}c). The finger-like protrusion grows towards the neighbouring channel side-wall (i.e. the path of least resistance) as it competes with the trailing bubble's tip. The trailing bubble's tip continues to advance at a constant speed whilst its front splits and, consequently, it maintains a constant separation from the middle bubble's rear. The invariance of the front's propagation speed whilst oscillating is similar behaviour to that which was reported in \citep{Pailha} and \citep{LifeAndFate}. Furthermore, the middle bubble's shape is preserved as the trailing bubble's front splits and, thus, it acts as a fixed perturbation to the flow field in the co-moving frame of reference of the trailing bubble's tip.


The splitting of the trailing bubble's front leads to the formation of complex patterns because it is not orthogonal to the direction of propagation and, thus, any disturbance is naturally advected away from its tip as it propagates \citep{curved_fronts}. The finger-like protrusion that develops underneath the trailing bubble's tip is advected towards its underside (i.e. where the normal component of the fluid's velocity vanishes) and it is left behind as the trailing bubble's tip advances. The protrusion remains near-stationary in the laboratory frame of reference and, thus, it appears to recede towards the trailing bubble's rear in the steadily advancing tip's frame of reference. The protrusion slowly relaxes before it is absorbed by the trailing bubble's rear because capillary pressure gradients drive flows that tend to minimise the perimeter of the air-liquid interface. The protrusions become increasingly flat as the distance from the trailing bubble's tip increases because the interface has relaxed for a longer period of time. Thus, although the deformation of the trailing bubble's front is perfectly periodic, the pattern that develops on its underside is not (figure~\ref{fig:three_bubble_oscs_Q106}).

\begin{figure}
\includegraphics[width=\textwidth, clip]{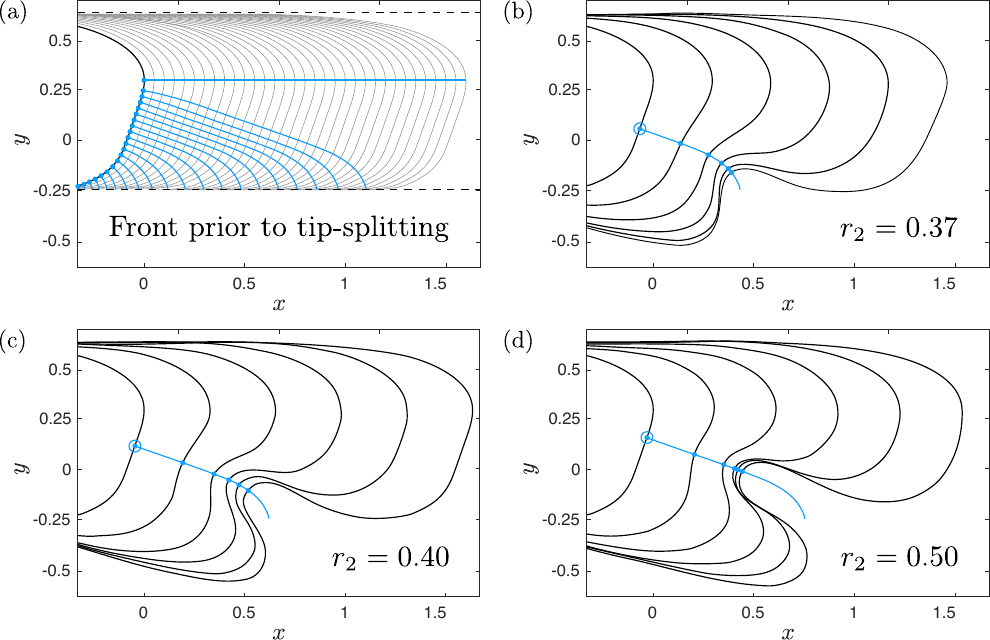}
\caption{(a) The trailing bubble's front instantaneously prior to splitting whilst the leading bubble's size $r_1 = 0.30$, trailing bubble's size $r_3 = 1.30$ and dimensionless flow rate $Q = \mu U_0^* / \sigma = 0.04$ are fixed. The grey-coloured curves are equally spaced translations of the front in the direction of propagation. The blue-coloured curves are instantaneously normal to the grey-coloured curves at their intersection points. (b--d) Composite images that show the spatiotemporal evolution of the trailing bubble's front during splitting as the middle bubble's size $r_2$ is varied. The normal curves that best-align with the indentations have been superimposed and the circular markers indicate the initial splitting position. The rail has been removed for visual clarity.}
\label{fig:normal_curves}
\end{figure} 

To explain the increase in the amplitude of the deposited pattern as the middle bubble's size is increased, we analyse the spatiotemporal evolution of the trailing bubble's front as it splits. The middle bubble's size does not influence the shape of the trailing bubble's front prior to splitting because its propagation speed, which is set exclusively by the leading bubble's size, is constant. The trailing bubbles front prior to splitting is presented in figure~\ref{fig:normal_curves}a and the grey-coloured curves correspond to a series of equally spaced translations of the front in the direction of propagation if it were to retain this shape. The blue-coloured curves originate at different points underneath the front's tip and they are instantaneously normal to the grey-coloured curves at their intersection points; they are referred to as the front's "normal curves" and correspond to the trajectories of liquid particles that are advected away from its tip. The motion of the indentation that develops on the trailing bubble's front arises exclusively due to its advection away from its tip and, thus, its trajectory corresponds to one of the front's normal curves \citep{curved_fronts}. We can utilise this property of the system --- by aligning the indentation's trajectory with one of the front's normal curves --- in order to precisely determine the initial splitting position \citep{lajeunesse}.

\begin{figure}
\includegraphics[width=\textwidth, clip]{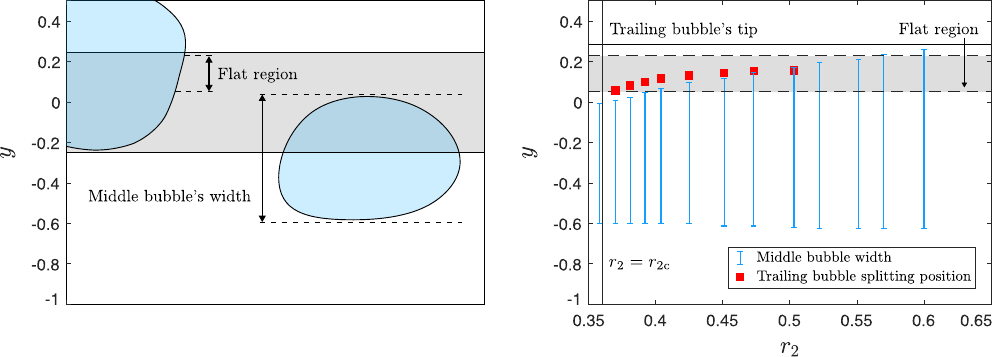}
\caption{The $y$-coordinates of the trailing bubble's initial splitting position and the edges of the middle bubble's cross section as functions of the middle bubble's size $r_2$. The leading bubble's size is $r_1 = 0.30$, the trailing bubble's size is $r_3 = 1.30$ and the dimensionless flow rate is $Q = \mu U_0^* / \sigma = 0.04$.}
\label{fig:middle_bubble_position}
\end{figure}

The composite images in figures~\ref{fig:normal_curves}(b--d) show the spatiotemporal evolution of the trailing bubble's front as it splits for various middle bubble sizes. We have superimposed the best-fitting normal curve to the indentation's trajectory in each case and the circular markers indicate the initial splitting position. The indentation originates closer to the trailing bubble's tip as the middle bubble's size is increased and, thus, the front splits increasingly symmetrically. The middle bubble's width increases with its size and, hence, it occupies a greater proportion of the channel's cross-section immediately ahead of the trailing bubble's front (figure~\ref{fig:middle_bubble_position}). The $y$-coordinate of the initial splitting position is always contained within the flat region (grey-coloured band) and it coincides with the maximum $y$-coordinate of the middle bubble's cross-section. The critical value $r_2 = r_{2\mathrm{c}}$ (i.e. the onset of oscillations) occurs when the maximum $y$-coordinate of middle bubble's cross-section crosses the centreline of the channel. This behaviour indicates that the middle bubble acts as an obstruction to the flow and, in doing so, it causes regions of the trailing bubble's front to move at different speeds. The region of the trailing bubble's front that is screened by the middle bubble advances slower than its tip and, thus, the front splits into two.

The transient growth of the finger-like protrusion is influenced directly by how the trailing bubble's front splits because it experiences different levels of inhibition as it competes with the trailing bubble's tip. This behaviour is largely similar to the growth of finger-like protrusions from an initially planar front: the fingers that are ahead tend to shield the others around them until a dominant finger wins. The trailing bubble's front is inclined and, thus, its tip shields the protrusion as it grows. This effect becomes more significant as the middle bubble's size is increased because the trailing bubble's front splits more asymmetrically (figure~\ref{fig:normal_curves}) and, thus, its tip starts increasingly further ahead. For example, the protrusion's growth is inhibited significantly for $r_2 = 0.37$ because the trailing bubble's front splits far away from its tip and, therefore, the quasiperiodic pattern that develops along its underside consists of stubby fingers (figure~\ref{fig:normal_curves}b). However, the protrusion's growth is inhibited less significantly for $r_2 = 0.50$ because the trailing bubble's front splits closer to its tip and, therefore, the quasiperiodic pattern that develops along its underside consists of pronounced fingers (figure~\ref{fig:normal_curves}d).

\subsection{Restoral dynamics}
\label{sec:restoral}

The middle bubble's shape does not change and the trailing bubble maintains a constant propagation speed as it oscillates. Thus, the trailing bubble's front has a preferred steady shape (i.e. the weakly unstable steady state from which the stable limit cycle emanates) that is set by the ratio of imposed viscous and capillary forces. The trailing bubble's tip broadens as it advances post-splitting because it is attracted to the weakly unstable steady state and, in doing so, it restores the initial shape that the front adopted prior to splitting. We are interested in how the dimensionless restoral time $T = 2 U_0^* T^* / W^*$ of the trailing bubble's front is affected by the system's parameters.

\begin{figure}[t]
\includegraphics[width=\textwidth, clip]{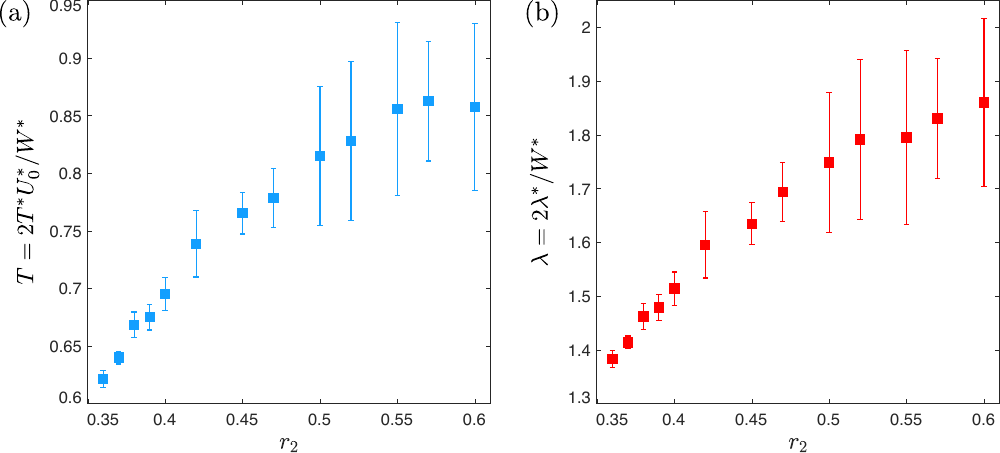}
\caption{Variation of (a) the front's dimensionless restoral time $T = 2 T ^* U_0^* / W^*$ and (b) the quasiperiodic pattern's dimensionless wavelength $\lambda = 2 \lambda^* / W^*$ as functions of the middle bubble's size $r_2$. The bubble sizes are $r_1=0.30$ and $r_3=1.30$ and the dimensionless flow rate is $Q = \mu U_0^* / \sigma = 0.04$. The error bars correspond to the standard deviations of the data sets.}
\label{fig:wavelength}
\end{figure} 

The dimensionless restoral time of the trailing bubble's front increases monotonically as the middle bubble's size is increased (figure~\ref{fig:wavelength}a). The balance of forces on the trailing bubble does not change as the middle bubble's size is increased because the trailing bubble's propagation speed is constant and, thus, we attribute this effect directly to the increasingly symmetrical splitting of its front. The front's tip becomes increasingly narrow post-splitting and, consequently, it is required to broaden more significantly before its initial (i.e. pre-splitting) shape is restored. The increase in the dimensionless restoral time of the trailing bubble's front is accompanied by a proportionate increase in the dimensionless spacing $\lambda = 2 \lambda^* / W^*$ between pairs of successive peaks in the quasiperiodic pattern that develops along its underside (figure~\ref{fig:wavelength}b) because the distance that the bubble advances between each instance of splitting increases by the same multiplicative factor. Thus, the frequency of the quasiperiodic pattern decreases as the front's restoral time increases at a fixed value of the bubble's propagation speed.

\begin{figure}
\includegraphics[width=\textwidth, clip]{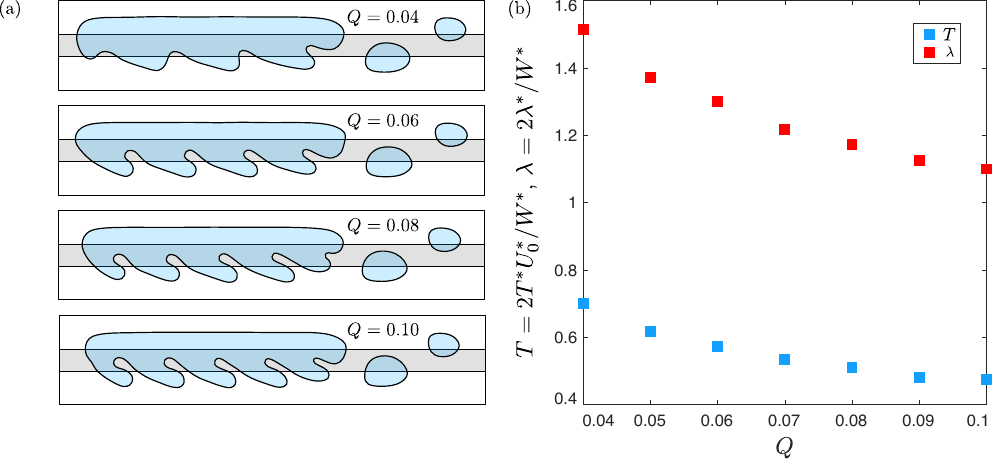}
\caption{(a) Snapshots of the patterns that are deposited by the trailing bubble as the dimensionless flow rate $Q = \mu U_0^* / \sigma $ is varied. (b) Variation of the front's dimensionless restoral time $T = 2 T ^* U_0^* / W^*$ and the pattern's dimensionless wavelength $\lambda = 2 \lambda^* / W^*$ as functions of the dimensionless flow rate $Q = \mu U_0^* / \sigma$. We have omitted error bars in (b) because they are smaller than the data markers. The bubble sizes $r_1=0.30$, $r_2=0.40$ and $r_3=1.30$ are fixed.}
\label{fig:lambdavaryQ}
\end{figure}

\begin{figure}
\includegraphics[width=\textwidth, clip]{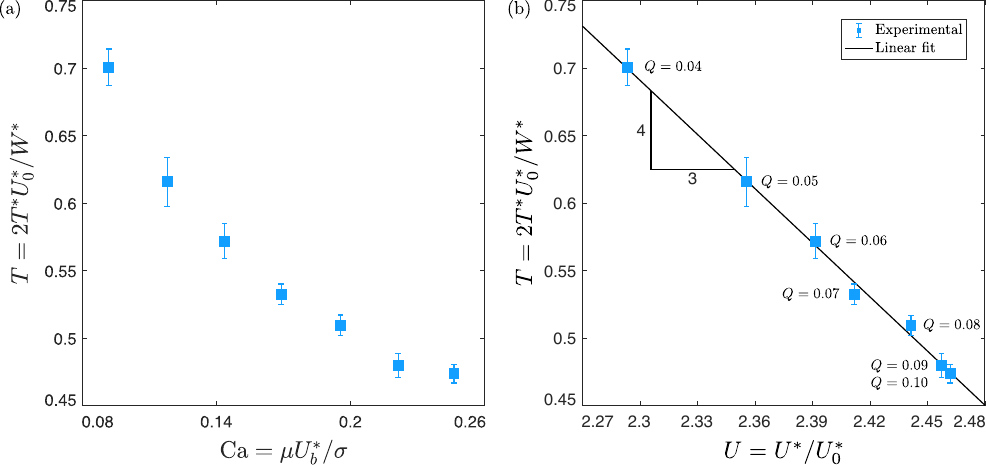}
\caption{Variation of the front's dimensionless restoral time $T = 2 T ^* U_0^* / W^*$ as functions of (a) the capillary number $\mathrm{Ca} = \mu U_b^* / \sigma$ and (b) the dimensionless bubble speed $U = U^* / U_0^*$.}
\label{fig:capillary_number}
\end{figure}

Finally, we investigate the influence of the dimensionless flow rate on the restoral dynamics of the trailing bubble's front whilst all of the bubble sizes are fixed. The frequency of the quasiperiodic pattern that develops along the trailing bubble's underside increases significantly as the dimensionless flow rate is increased (figure~\ref{fig:lambdavaryQ}a). This behaviour is because the dimensionless restoral time of the trailing bubble's front decreases monotonically as the dimensionless flow rate is increased (figure~\ref{fig:lambdavaryQ}b). The trailing bubble's propagation speed increases as the dimensionless flow rate is increased and, therefore, the influence of viscous forces that arise from its propagation through the surrounding liquid increases relative to the constant surface tension. The quasiperiodic pattern is, in fact, nearly perfectly periodic at the highest flow rates; this indicates that the capillary pressure-induced restoral of the trailing bubble's interface, which would act to damp the oscillations, is negligible. Thus, the restoral timescale of the trailing bubble's front is set predominantly by viscous forces over the investigated range of parameters --- in the same manner as the bubble's dimensionless velocity $U = U^* / U_0^*$. The increasing influence of viscous forces with flow rate enhances the rate at which the trailing bubble's tip readjusts post-splitting and, consequently, the front's restoral time decreases with the capillary number (figure~\ref{fig:capillary_number}a). This interpretation is further supported by the linear decrease of the front's dimensionless restoral time with its dimensionless propagation speed (figure~\ref{fig:capillary_number}b).


\section{Conclusion}
\label{conclusion}

We have investigated the dynamics of stable periodic propagation modes that arise in the propagation of a group of air bubbles in a Hele-Shaw channel that contains an axially uniform depth reduction (rail) along its centreline. The study is motivated by a desire to understand the physics that govern periodic dynamics in finger propagation because the subcritical transition from steady to disordered propagation may be underpinned by weakly unstable periodic states. The system under investigation readily exhibits periodic dynamics and, thus, it is a suitable playground to explore the complex nonlinear behaviour of interest.

The local reduction in the channel's depth that is induced by the rail allows groups of bubbles to organise into stable spatial arrangements because the local changes in viscous resistance provide a mechanism to facilitate changes in their shapes and, hence, propagation speeds. We exploited this particular property of the system by using it as a method of remotely applying a sustained perturbation to a bubble's shape as it propagates. The system undergoes a supercritical Hopf bifurcation for particular combinations of its parameters and the bubble's front deforms periodically as the system explores the emergent stable limit cycle. We identified a generic physical mechanism that leads to periodic dynamics in this system by splitting the front's dynamics during a single oscillatory cycle into two fundamental physical components: splitting and restoral. The interplay between these two components causes the bubble to deposit a quasiperiodic pattern of finger-like protrusions along its underside as it propagates. We related the qualitative features of the quasiperiodic pattern to the bubble's splitting and restoral dynamics following variation of the system's parameters. The protruding fingers increase in amplitude as the bubble's front splits closer to its tip because they are inhibited less. The dimensionless restoral time of the oscillating front decreases as the capillary number increases and this leads to an increase in the frequency of the quasiperiodic pattern. The capillary pressure-induced restoral of the interface is negligible within our investigated parameter regime and, thus, the timescale of restoral is set predominantly by viscous forces that arise from the front's propagation through the surrounding liquid.

Importantly, unlike previously observed oscillations during finger propagation in rail-occluded Hele-Shaw channels (e.g. \citep{Pailha}), the generic mechanism that was identified in this paper does not directly involve the geometry of the rail and this means that it could be present in a Hele-Shaw channel without a rail. We have established that perturbations can, therefore, lead to periodic dynamics in this system: an insight that will be valuable in future studies of the subcritical transition from steady to disordered propagation. However, it remains an open question as to whether or not disordered propagation is, indeed, underpinned by a multitude of weakly unstable periodic states.

\acknowledgements{This work was supported via EPSRC grants EP/P026044/1 and EP/T008725/1 and an EPSRC DTP studentship (JL).}

\bibliography{bibliography}

\end{document}